# What does the tree of life look like as it grows?
## Evolution and the multifractality of time

**Kevin Hudnall (corresponding author)**

kahudnall@ucdavis.edu

Biological Systems Engineering Graduate Group

University of California, Davis, CA 95616 USA

**Raissa M. D'Souza**

rmdsouza@ucdavis.edu

University of California, Davis, CA 95616 USA and
Santa Fe Institute, Santa Fe, NM, 87501, USA

## Highlights

- We develop a mathematical formalism to model the dynamic tree of life based on three foundational principles of biology: the nestedness, duality, and randomness of phylogeny.
- The resulting structure resembles a Cantor dust, with each branch being a distinct fractal curve, so that the living tree of life is multifractal in the sense that it consists of many unique fractals.
- Since branch lengths represent time intervals, the length of any time interval will depend on the location of the observer in the system (i.e., its biological form), so that time is also multifractal.

## Abstract

By unifying three foundational principles of modern biology, we develop a mathematical framework to analyze the growing tree of life. Contrary to the static case, where the analogy between phylogenetic trees and the tree that grows in soil is drawn, our framework shows that the living tree of life is analogous to a Cantor dust where each branch is a distinct fractal curve. The system as a whole is therefore multifractal in the sense that it consists of many unique fractals. The three foundational principles for the mathematical framework are that phylogeny is nested, phylogeny is dualistic (i.e., transitive between singularities and populations), and phylogeny is stochastic. Integrating these three principles, we model the dynamic (i.e., *living*) tree of life as a random iterated function system that generates unique convexly related sequences of branching random variables (visualized in **Animation 1**). The multifractal nature of this dynamic tree of life implies that, for any two living entities, the time interval from their last common ancestor to the present moment is a distinct fractal curve for each. Thus, the length of a time interval along each distinct branch is unique, so that time is also multifractal and not an ultrametric on the tree of life.

## 1 Introduction

The tree that grows in soil has informed evolutionary biology throughout its history (Archibald, 2009). But how is the tree of life fundamentally different than the tree that grows in soil? When the tree in soil grows a layer, all previous layers back to the trunk continue to exist. Not so with the tree of life. As the tree of life grows a layer, only the leaves (i.e., the living) exist, and everything back to the trunk (i.e., the dead) does not. The living tree of life consists of sequentially realized, then eliminated biological forms: the living beget the living and disappear into the nonliving. Descent-with-modification means these sequences must be "nested" and "branching". The stochasticity of evolution means these sequences must be sequences of random variables.

The objective of this work is to develop a mathematical framework that brings together these foundational principles to capture the structure of the living tree of life. In doing so, we derive a structure that represents the living tree of life as a "randomly branching dust". This brings a dynamical perspective to phylogenetics, which has previously focused primarily on static representations of the tree of life.

In the next section we explain the three principles of modern biology which form the cornerstones of our mathematical formalism: phylogeny is "nested" (**the nested principle**), phylogeny is "dualistic" (i.e., transitive between singularities and populations) (**the duality principle**), and phylogeny is stochastic (**the random principle**). Then in Section 3 we show how these three principles are modeled mathematically as a random iterated function system, followed by Section 4 which presents simulation results and demonstrates the multifractality of the structure. In Section 5 we discuss the strengths of the model as well as the implications of the multifractal structure, and finally Section 6 offers some concluding remarks.

## 2. Foundational biological principles

The three principles of nestedness, duality, and randomness are foundational to modern biology and can each be clearly stated mathematically. We begin with the nested principle.

### 2.1 The nested principle

"Nestedness" is a consequence of the logical structure of descent-with-modification: species trees are "contained within" genera trees, which are "contained within" family trees, and so on ("groups subordinate to groups" in Darwin's words (1859)). As part of the inherent structure of descent-with-modification, "nestedness" has been central to mathematical models of evolution from the earliest to the most recent (Ane et al., 2017; Doulcier et al., 2020; Fontana & Buss, 1994; Ponisio et al., 2019; Semple, 2016; Yule, 1925).

"Nestedness" can be made precise using the convexity inherent in ancestor-descendant relationships: the probability of the child is always less than the probability of the parent since the parent is a precondition of the child's existence. The "nestings" are then sequences of "intersecting ancestors". Mathematically, these sequences of "intersecting ancestors" are sequences of intersecting sets, where the sets being intersected are the supports of the random variables. Due to random generation, these sequences form unique convex hulls through the tree of life which we call "paths" (Definition A1, Appendix A.1). (The convex hull of a set is the intersection of all sets that contain it.)

*Nested sequence notation*

We use the Greek $\beta$ to stand for a biological object. Nested subscript notation is used to designate the unique paths through the tree of life. To represent "species" $k$ of "genus" $j$ of "family" $i$, we write $\beta_{i_{j_k}}$. To represent the (sub)tree generated by $\beta_{i_j}$ we write $\mathcal{T}_{i_j}$. **Figure 1, Row I** illustrates the notation and the convexity of phylogenetic space.

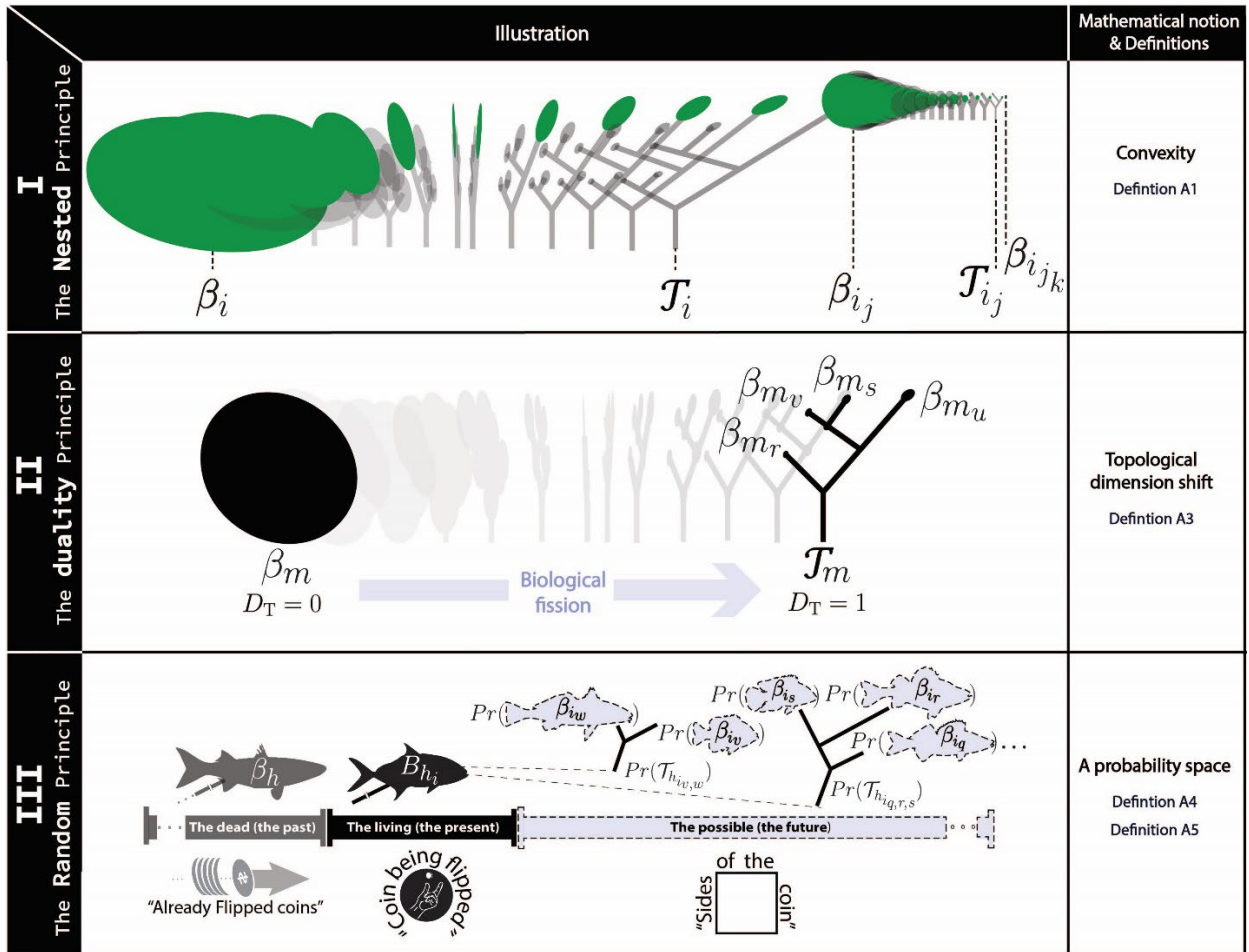

**Figure 1** *Three foundational principles of modern biology that suggest the multifractality of time.* **(ROW I)** *The nested principle.* The path leading to "species" $k$ (denoted $\beta_{i_{j_k}}$) is a unique sequence of convexly nested sets where the topological dimension stays constant at 0 (the dimension of a point). The convex hull of "species" $k$ is the intersection of all its ancestors. This intersection forms a path (highlighted green). **(ROW II)** *The duality principle.* Transformation of biological form is a transition from topological dimension zero ($D_T = 0$) to topological dimension one ($D_T = 1$). $\mathcal{T}_m$ is the downscale subtree generated by the upscale parent leaf $\beta_m$. $\mathcal{T}_m$ consists of the distances separating the set of leaves $\{\beta_{m_r}, \beta_{m_v}, \beta_{m_s}, \beta_{m_u}\}$. Phylogenetic space is dualistic in the sense that it is composed of such fissions. **(ROW III)** *The random principle.* $B_{h_i}$ is alive: it is a random variable with a nonzero probability of at least one possible outcome (it is "the coin being flipped"). Possible outcomes of $B_{h_i}$ (the "sides of the coin") are given at the right with dashed outlines as two trees of descendants. Each member fish of the tree has a particular probability of being realized by $B_{h_i}$, as does the (sub)tree itself. $\beta_h$ is the immediate ancestor of $B_{h_i}$, with the convex hull of $B_{h_i}$ being the intersection of the supports of all its ancestors. The members of the chain of "flipped coins" from the ancestor of all life to entity $h$, have no events with a nonzero probability of occurring and so they are dead. Note that we use the capital $B$ (with appropriate subscript) to represent the random variable, and the lower case $\beta$ (with appropriate subscript) to represent an event.

## 2.2 The duality principle

The second foundational principle is the "duality" of phylogenetic space: the tree of life consists of transitions between singularities and populations. This too is a central tenet of Descent-with-modification, and as such there is a large body of literature on such "evolutionary transitions" (Eigen, 2000; Fontana & Schuster, 1998; Jablonka, 1994; McShea, 2001; Shelton & Michod, 2014; Szathmáry & Smith, 1995). Diversification from a common origin means these transitions are multiplicative; and we will follow (Chou & Greenman, 2016; Grey et al., 1995) in calling them "fissions". "Fission" models abound as Yule processes (Gernhard et al., 2008; A. Mooers et al., 2012), birth-death processes (Chou &

Greenman, 2016; Hallinan, 2012), Brownian motion processes (Czuppon & Gokhale, 2018), and branching processes (Bellman & Harris, 1949; Grey et al., 1995; Kimmel & Axelrod, 2015; Loeffler & Grossmann, 1991).

The duality principle allows us to model each fission event as a transition from a point (i.e., leaf) to a tree with the topological dimension jumping from zero to one. Along with these dimension shifts, a basic convexity still holds: a "species" tree is contained within a "genus" leaf in the sense that the support of each "species" leaf is a convex subset of the support of the "genus" leaf. Using this convexity, and the shifts in topological dimension, we define these "nested fissions", and thereby arrive at a definition of "duality" (Definition A3, Appendix A.2). **Figure 1, Row II** illustrates these transitions.

Evolutionary transitions also occur as "fusions", where distinct extant lineages merge to form new objects. Sexual reproduction and multicellularity are instances of biological "fusion". However, comparing distantly related lineages on the tree of life (i.e., at a global scale), fission will be the dominant force. This current work deals only with fission. Incorporating "fusion" would require shifting from a strictly tree representation to a general network representation. Additionally, fission is antecedent to "fusion" since branches must first exist if they are to merge, so it is natural to start with fission.

## 2.3 The random principle

The third foundational principle of modern biology, that phylogeny is stochastic, is the one of which we will have the least to say. The notion of randomness is at the center of modern biology, with the ultimate source of biological variability believed to be random genetic mutations. To Dobzhansky's famous quote that "nothing in biology makes sense except in the light of evolution" (1973), we might add that nothing in evolution makes sense except in the light of random generation. Nearly all modern biological works employ randomness.

As one traverses a path in the tree of life, one traverses a long chain of the dead. In the limit over this chain of the dead, one arrives at the living – the random variable. The probability of the living depends on the long chain of dead ancestors that constitute that particular living entity's unique convex hull. **Figure 1, Row III** illustrates the randomness of phylogenetic space.

## 2.4 Connecting the three principles

In broad outline, here is how the three principles connect to model the living tree of life. The random principle allows the tree of life to be modeled as a probability space. This in turn allows the nested principle and the duality principle to be mathematically formulated using probability theory. It also allows life and death to be formulated in probabilistic terms: entities that have a nonzero probability of fission are alive; entities that have zero probability of fission are dead (Definitions A4, A5, Appendix A.3). Every lineage in the tree of life is a unique chain of random variables where the probability of the descendant depends on the ancestor. To strip away biological complexity, we use uniform distributions (see Section 3.1.2). The nested principle means the support of the descendant is a subset of the support of the ancestor. The duality principle means these sequences of random nested subsets are branching, and therefore unique convex hulls are "carved out" through the tree of life. The convexity further means that in the large reproduction limit (Section 3.3) over any particular path, there is approximately a single entity at the "end" of each chain. These unique entities at the "ends" of the chains are the living: random variables with a nonzero probability of becoming something else (i.e., fission).

## 3 Materials and Methods

In this section we show how the three principles of nestedness, duality, and randomness are modeled mathematically as a random iterated function system. We start by developing the probability space, then proceed to give the iterated function system, followed by an analysis of convergence, and finally an analysis of the fractal dimensions of the paths generated.

### 3.1 The tree of life as a probability space

There are two kinds of objects, modeled as points and lines respectively: **leaves** of topological dimension zero (denoted $\mathbb{L}^{D_T=0}$); and **trees** of topological dimension one (denoted $\mathbb{T}^{D_T=1}$). Their respective symbols are $B$ and $\mathcal{T}$ (along with their proper subscript identifiers). The motivation for this choice of $D_T = 0$ and $D_T = 1$ is mostly mathematical. To represent time one needs distances ($D_T = 1$) and thus trees are used to capture multiplicity. To represent the living, one needs a notion of singularity or "individuality" and thus points ($D_T = 0$) are used to capture the living things at the "ends" of the lineages (Section 3.3).

The random principle means phylogenetic space can be modeled as a probability space:

**Expression 1**
$$(\mathbb{L}^{D_T=0} \cup \mathbb{T}^{D_T=1}, \mathcal{F}, Pr).$$

Here $\mathbb{L}^{D_T=0} \cup \mathbb{T}^{D_T=1}$ is the set of all possible biological forms (i.e., leaves of topological dimension zero and trees of topological dimension one); $\mathcal{F}$ is a σ-field; and $Pr$ is a probability measure.

#### 3.1.1 Probability functions, life, and death

The living entities are the random variables. We use the upper-case beta "$B$" to represent random variables; and the lower-case beta "$\beta$" to represent events. To represent the unique chains of random variables, we use subscript notation. We use a colon ":" to represent an unspecified gap in a lineage, and zero "0" to represent the ancestor of all life. So to specify the path from the ancestor of all life to "species" $k$ of "genus" $j$ of "family" $i$, we write:

**Expression 2**
$$B_{0_{:i_{j_k}}}.$$

Often this will just be shortened to $B_{:k}$ or $B_k$, and the sequence will be implied; but the notation used in Expression 2 is useful both conceptually, and when taking sequence limits.

A continuous random variable $B_{:m}$ has a **distribution function**:

**Equation 1**
$$F_{:m}(x) = Pr(B_{:m} \leq x), \qquad x \in supp(B_{:m}),$$

where $supp(\cdot)$ is the support of the random variable's density function. The subscript on $F$ has a dual purpose: it indicates the random variable under consideration ($B_{:m}$ in this case); it also reminds us that we are dealing with a nested sequence of distribution functions.

The distinction between life and death can be stated using these distribution functions. A biological object $B_m$ is **alive** if there exists an $x$ for which $F_m(x) \neq \{0,1\}$. Likewise, a biological object $B_m$ is **dead** if there does not exist such an $x$. See Appendix A.3 for more on the definitions of life and death.

#### 3.1.2 Probability space implementation

We implement the model by choosing uniform probability distributions for the random variables. The use of uniform distributions means the model is one of drift, and so selection is not included. This is however the correct starting point since the existence of drift means selection is not a necessary condition of evolution. Further, the implementation assumes all entities reproduce once and for all at the end of their lives. Without this assumption (i.e., if parent lives overlap child lives), one needs notions of partial convexity.

Nestedness is obtained by making biological entity $m$ a random variable distributed uniformly from zero to the form of its most recent ancestor:

**Expression 3**

$$B_{0:l_m} \sim \text{Uniform}\left(0, \beta_{0:l}\right) \text{ with } \beta_0 = 1.$$

Here the initial condition (the ancestor of all life) is equal to 1 (i.e., $Pr(B_0 = 1 = \beta_0) = 1$). The support of $B_{:m}$ is then given as: $supp(B_{:m}) = [0, A(B_{:m})]$, where $A(\cdot)$ is a function that returns the most recent ancestor of its argument. Expression 3 further means that the Markov property is met, and each descendant only depends on the probability of its most recent ancestor:

**Equation 2**

$$Pr\left(B_{0:k_{l_m}} \leq x \middle| B_{0:k_l} = \beta_{0:k_l}, B_{0:k} = \beta_{0:k}, \ldots, B_0 = \beta_0\right) = Pr\left(B_{0:k_{l_m}} \leq x \middle| B_{0:k_l} = \beta_{0:k_l}\right).$$

Under Expression 3, the supports of the random variables are strictly decreasing orders on the unit interval:

$$[0, \beta_0] \supset \cdots \supset \left[0, \beta_{0:j}\right] \supset \left[0, \beta_{0:j_k}\right] \supset \left[0, \beta_{0:j_{k_l}}\right].$$

Expression 3 also means that the **distribution function** $F$ is given as:

**Equation 3**

$$F_{:l_m}(x) \equiv Pr\left(B_{:l_m} \leq x\right) = \frac{x}{\beta_{:l}}, \quad x \in [0, \beta_{:l}),$$

and the **density function** $f$ takes the form:

**Equation 4**

$$f_{:l_m}(x) = \frac{1}{\beta_{:l}}, \quad x \in [0, \beta_{:l}).$$

The subscript on $f$ is a slight abuse of notation. The sequence of functions $\{f_{:l_m}\}$ is not a nested sequence. Instead, $\{f_{:l_m}\}$ is a strictly increasing sequence that goes to infinity as $\beta_{:m}$ goes to zero.

**Figure 2** illustrates the pathwise sequences of distribution and density functions.

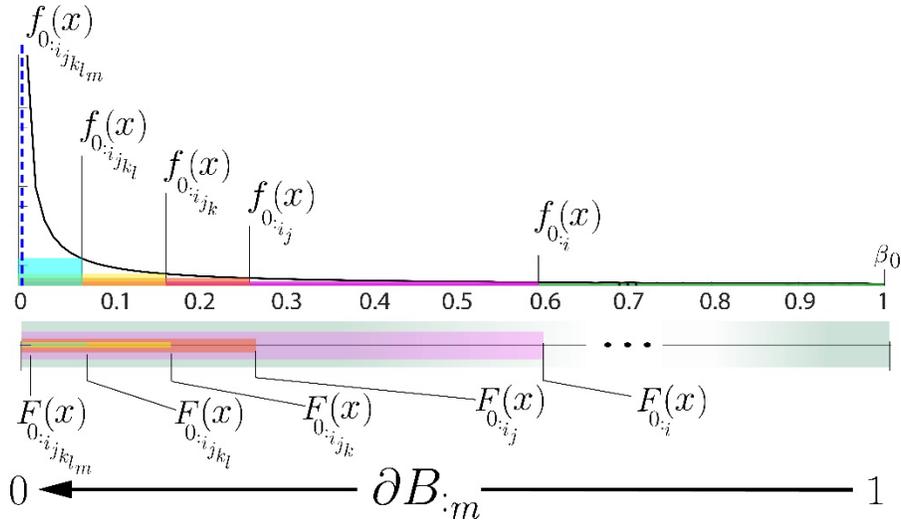

**Figure 2** *Paths as sequences of probability functions that generate unique unit interval partitions.* Differential form of path $m$, denoted $\partial B_{:m}$, is modeled as the path-specific limit over strictly decreasing sets of nested probability distribution functions $\{F_{:m}\}$. This means the set of probability density functions $\{f_{:m}\}$ for any path $m$ is then a unique strictly increasing sequence. By this formulation, every path through the tree of life forms a unique sequence of unit interval partitions that are nested as defined in Definition A1, Appendix A.1.

Since leaves are the limits over nested sequences of distribution functions, the diameter of a living biological object is taken to be the supremum of its support, which in our uniform implementation is simply its most recent ancestor:

**Definition 1**

$$diam\left(B_{:l_m}\right) \equiv sup\ supp\left(f_{:l_m}\right) = A\left(B_{:l_m}\right) = \beta_{:l}.$$

## 3.2 The random iterated function system

In this section we implement the principles of nestedness, duality, and randomness as an iterated function system. The iterated function system generates a random tree, and then for every leaf of the generated tree, it generates another random tree at a scale strictly less than the parent tree. The structure is mapped to the $n$th-dimensional unit interval $[0,1]^n$, where $n$ is the number of reproductions.

*Auxiliary function*

To specify the iterated function system, we first need a simple auxiliary function that returns the number of leaves (terminal nodes) in any subtree. Let $L: [0,1] \rightarrow \mathbb{Z}^+$ such that $L\left(\mathcal{T}_{0_{:i}}\right) =$ the number of leaves of tree $\mathcal{T}_{0_{:i}}$.

*Ratio lists*

The ratio lists represent lists of leaves. Each member of a ratio list gives the scale at which a random tree is generated. Every leaf in a ratio list has generated within it a scaled-down ratio list, every member of which has generated within it another scaled-down ratio list, and the process is carried out iteratively in this manner. The dependence of the "child" tree on the "parent" leaf is accomplished by making scale variables (i.e., members of the ratio lists) distributed over nested intervals of the unit interval using Expression 3 of Section 3.1.2.

For tree $\mathcal{T}_{0_{:i}}$ in the $K^{\text{th}}$ system iterate, its ratio list in the $(k+1)^{\text{th}}$ iterate is given as:

Expression 4$$\left\{\beta_{0:i_1}, \beta_{0:i_2}, \ldots, \beta_{0:i_{L(\mathcal{T}_{0:i})}}\right\}.$$

In Expression 4, $i$ is the parent leaf and the set of leaves $\left\{\beta_{0:i_1}, \beta_{0:i_2}, \ldots, \beta_{0:i_{L(\mathcal{T}_{0:i})}}\right\}$ are the leaves of the tree generated by $i$. As a simplification we impose: $\beta_{0:i_1} = \beta_{0:i_2} = \cdots = \beta_{0:i_{L(\mathcal{T}_{0:i})}}$. That is, every leaf of a particular generated subtree is at the same scale.

If the $K^{\text{th}}$ system iterate contains additional subtrees other than $\mathcal{T}_{0:i}$, then each one of those subtrees will correspond to a ratio list in the $(k+1)^{\text{th}}$ iterate that has as many members as there are leaves in the subtree in the $K^{\text{th}}$ iterate. So the $(k+1)^{\text{th}}$ iterate will have as many ratio lists as there are subtrees in the $K^{\text{th}}$ iterate.

*Function lists*

The function lists correspond to randomly generated trees at scales of the ratio lists. The leaves of tree $\mathcal{T}_{0:i}$ in the $K^{\text{th}}$ system iterate, are comprised of $L(\mathcal{T}_{0:i})$ executions of function $T^K_{0:i_j}$, so that the function list of the $(k+1)^{\text{th}}$ iterate is:

Expression 5

$$\left\{T^K_{0:i_1}, T^K_{0:i_2}, \ldots, T^K_{0:i_{L(\mathcal{T}_{0:i})}}\right\}.$$

Here, the iterated function $T^K_{0:i_j}$, $j = 1$ to $L(\mathcal{T}_{0:i})$, takes as its argument the corresponding ratio list member (the scale of the parent leaf), and generates the new child tree $\mathcal{T}_{0:i_j}$:

Equation 5

$$T^K_{0:i_j}\left(\beta_{0:i_j}\right) = \{Z\} * \beta_{0:i_j} = \mathcal{T}_{0:i_j}.$$

Here $\{Z\}$ is a sequence generated by the Galton-Watson branching process:

Equation 6

$$Z_{p+1} = \begin{cases} \xi_1^{p+1} + \cdots + \xi_{Z_p}^{p+1}, & Z_p > 0 \\ 0, & Z_p = 0 \end{cases},$$

where the $\xi_r^p \in \mathbb{Z}$ are $i.i.d.$ nonnegative random variables. For the variables $p$ and $r$ we have: $0 < p < G$, and $0 \leq r < M$. Here $M$ gives the maximum number of offspring possible and $G$ gives the maximum number of generations possible.

If the $K^{\text{th}}$ iterate contains additional subtrees other than $\mathcal{T}_{0:i}$, then they too will have an associated function list. Each tree in the $K^{\text{th}}$ iterate will have a corresponding function list in the $(k+1)^{\text{th}}$ iterate; and each function list in the $(k+1)^{\text{th}}$ iterate will have as many members as there are leaves in its corresponding subtree in the $K^{\text{th}}$ iterate.

Equation 5 generates a random tree and randomly scales it at a value less than its parent's scale. The upper bounds on $p$ and $r$ force all Galton-Watson trees generated to be finite, and so questions of criticality do not arise. Indeed, it is unclear how one would embed trees in leaves that grow on infinitely long branches; so taking all biological beings as finite seems an important step. The results in this work were generated setting $G$ and $M$ equal to 2 and 3 respectively.

The iterated function system generates the tree of life so that every lineage is a sequence of decreasing nested random variables that uniquely partition the unit interval. These sequences are paths as in Definition A1, Appendix A.1. The paths form groups of subtrees so that as we change scale ($\partial B_{:m}$), fission as in Definition A3, Appendix A.2 occurs, and we shift from leaves (points) to trees (lines). Appendix B offers an illustrative example of the iterated function system.

### 3.2.1 Visualization on the unit square

If we wish to visualize the structure on the unit square (as in Animation 1), then we need two additional auxiliary functions: one that returns the coordinates of the root of any subtree (denoted $R$), and one that returns the coordinates of leaves within a subtree (denoted $l$). Let $R: [0,1]^2 \to [0,1]^2$ such that $R\left(\mathcal{T}_{0_{:i}}\right) = (a,b)$, the coordinates of the root node of tree $\mathcal{T}_{0_{:i}}$. And let $l: [0,1]^2 \to [0,1]^2$ such that $l\left(\mathcal{T}_{0_{:i}}\right) = \{(p,q), (r,s), \ldots\} \equiv \left\{l_{0_{:i_1}}, l_{0_{:i_2}}, \ldots\right\}$, the coordinates of the leaves of tree $\mathcal{T}_{0_{:i}}$. Then Equation 5 becomes:

**Equation 7**

$$T^K_{0_{:i_j}} = \{\mathbf{Z}\} * \beta_{0_{:i_j}} + ||R\left(\mathcal{T}_{0_{:i_j}}\right) - l_{0_{:i_j}}||_2, \qquad l_{0_{:i_j}} \in \mathcal{T}_{0_{:i}}$$

Here the term $||R\left(\mathcal{T}_{0_{:i_j}}\right) - l_{0_{:i_j}}||_2$ shifts the newly generated tree $\mathcal{T}_{0_{:i_j}}$ so it occupies the same location on the unit square as its parent leaf $l_{0_{:i_j}}$.

### 3.3 Cantor's nested set theorem

Since leaves are the limits over unique Markov chains of random variables, convergence questions naturally arise. A theorem from topology/analysis guarantees that the limit of each path through the tree of life contains a single entity. Since the supports of the chains of random variables are strictly decreasing under Expression 3, the sequence limit of a path's diameter (Definition 1) goes to zero as the path length goes to infinity (i.e., as reproductions $n$ goes to infinity):

**Equation 8**

$$\lim_{n \to \infty} diam(B_{:m}) = 0.$$

We therefore have met the hypotheses of **Cantor's nested set theorem** (Cohn, 2013) stated as follows:

**Theorem 1**

Let $supp(B_{:m}) = \left\{\bigcap_{i=0}^{\infty} [0, \beta_{:i_:}]\right\}$ be a decreasing sequence of nonempty closed sets on $[0,1]$, with $\lim_{n \to \infty} diam(B_{:m}) = 0$. Then $supp(B_{:m})$ contains exactly one point.

In the infinite limit, the point at the end of the chain is zero itself: every path converges in probability to zero since for all $\epsilon > 0$

**Equation 9**

$$\lim_{n \to \infty} Pr[|B_{:m} - 0| > \epsilon] = 0.$$

This is reasonable biologically since death is at the limit of every biological path. However, while the path is alive/extant, the limit at infinity is never reached, so that in the living case, each path becomes an approximation to the limit at infinity. If the tree of life is very old (i.e., reproductions $n$ is large), the living entities become large iteration (denoted $lim_{n \gg 1}$) approximated "ends" of the paths. See Section 5.1 for a discussion.

### 3.3.1 From the dead to the living, sequences of $\sigma$-fields

The nested principle, duality principle, and random principle imply that the tree of life consists of nested sequences of branching convex random variables indexed by path. This means that the dead tree of life $\mathbb{T}^{\text{dead}}$ consists of sequences of σ-fields, and therefore also requires an index: $\mathbb{T}_n^{\text{dead}}$, where $n$ gives reproductions. Under the approximation of Cantor's nested set theorem, the living tree of life is then the sequence limit of the dead tree of life:

**Equation 10**
$$\mathbb{T}^{\text{alive}} = lim_{n \gg 1} \mathbb{T}_n^{\text{dead}}.$$

## 3.4 Multifractal detrended fluctuation analysis

The tree of life generated by the iterated function system constitutes a multifractal. To demonstrate this (Section 4.3) we use multifractal detrended fluctuation analysis to calculate generalized Hurst indices (Kantelhardt et al., 2002). First, the cumulative sum of the deviations from the mean is calculated to obtain a transformed series for each path. These transformed series are then segmented into non-overlapping boxes of varying sizes (the sizes used for the data in this work were 3, 4, 5, 6, and 7). A linear fit is then applied to the data over each box, and residuals are calculated. These residuals are used to calculate a moment-dependent fluctuation function $V_q(n)$, where $q$ gives the moment and $n$ gives the number of points. The values for $q$ used in this work were $-5, -3, -1, 0, 1, 2, 3$, and $5$. A regression on the log of the fluctuation function versus the log of the box size is then performed, and the moment-dependent generalized Hurst exponent is given by the slope of the regression line.

This procedure is performed for each path in the system. If the generalized Hurst exponent is constant across different moments $q$, it indicates that the path is monofractal. If each path has a distinct generalized Hurst index, it indicates that the system as a whole is multifractal in the sense that it consists of many distinct monofractals.

## 3.5 The correlation dimension as the fractal dimension

Here we show how the fractal dimensions of the paths are calculated. These paths are sequences of random points where the clustering of the points on the unit interval varies between paths. We will therefore use the correlation dimension (Grassberger & Procaccia, 1983) to represent the fractal dimension of a path. The correlation integral is given as:

**Definition 2**
$$C(\varepsilon) = \lim_{N \to \infty} \frac{g}{N^2},$$

where $g$ gives the number of pairs of points whose difference is less than ε, and $N$ gives the total number of points, which in this case is the same as the number of iterations. The correlation integral scales as:

**Expression 6**
$$C(\varepsilon) \sim \varepsilon^{D_F},$$

where $D_F$ is the correlation dimension (denoted $D_F$ for "fractal dimension").

### 3.5.1 Numerical determination of the correlation dimensions $D_F$

When $N$ is sufficiently large, the correlation dimension can be estimated as the slope of the log of the correlation integral versus the log of ε. To find the correlation dimension of a path we first specify a one dimensional array of ε values. These values decrease logarithmically from one to zero, but do not include the endpoints one and zero. It is however necessary that the array contain values close enough to zero to be less than the minimal distance between any two points in the path. At each value of ε, the number of pairs of points in the path with a distance less than ε are determined and divided by the square of the total number of points in the path $N$, so that we have an array $C(\varepsilon)$ that is the same size as ε. Once we arrive at a value of ε for which $C(\varepsilon)$ is zero, we stop and exclude this last value. The log of ε is regressed against the log of $C(\varepsilon)$, and the slope of the regression line gives $D_F$. See Appendix C for more on this method.

## 3.6 Assumptions of the model

Several simplifying assumptions are built into the model, and we gather them here for clarity and convenience. First, the lack of fusion (Section 2.2) means this is a model of asexual reproduction only. Incorporating fusion requires shifting from a strictly tree representation to a general network representation so that transitions occur between topological dimensions zero, one, and now two as well. Second, entities reproduce once and only once at the end of their lives, so that ancestors' lives do not overlap descendants' lives. The main challenge in removing this assumption is developing the mathematics of partial convexity such a removal requires. Third, our current implementation of the model using uniform probability distributions (Section 3.1.2, Expression 3), as well as holding the maximum possible number of generations and offspring constant in the branching process (Section 3.2) means this implementation is one of drift only, and selection is not included. We contend however, that this particular assumption is the correct starting point for developing biological theory since the existence of drift means selection is not a necessary condition of evolution. Additionally, the use of only uniform probability distributions also means that the model employs a very limited notion of inheritance. Fourth, every leaf in a subtree is at the same scale (Section 3.2). This assumption can be removed relatively easily by randomizing scale within a subtree, though doing so does not yield additional insight. Finally, we note that an extension of the model making it more biologically realistic would be to use a continuous branching process instead of discrete branching processes (Equation 6).

It is our expectation that the removal of these assumptions will not compromise the main finding in this report, the multifractality of time, but instead enhance it. That is, we expect that lifting these assumptions will result in a greater diversity of lineages in the tree of life and therefore a greater variability in scaling behavior amongst the lineages resulting in more severe multifractality. That multifractality exists with the minimum evolutionary force should assure us it will exist when evolution is modeled in full force.

# 4 Results

Here we present simulation results for the iterated function system, give a brief preview of how information theory is applied to the structure, demonstrate that the structure is multifractal, and give simulation results for the fractal dimensions.

## 4.1 The living tree of life is a dust

**Animation 1** presents a realization of the stochastic iterated function system mapped to the unit square.

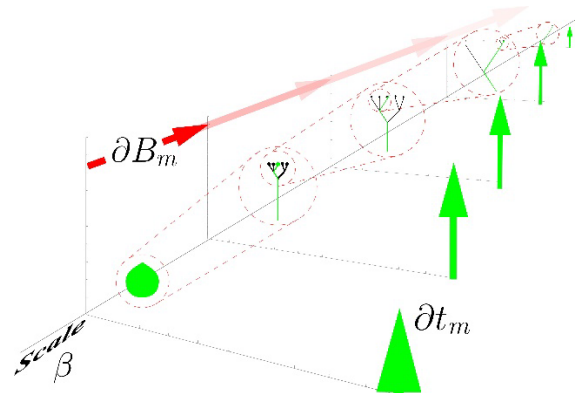

**Animation 1** *A model of the living tree of life.* **(LEFT)** A living biological entity $B_m$ is a path-specific random variable assigned to scale. The differential biological form $\partial B_m$ is then a contraction of scale along path $m$, by which a random subtree is generated at a random scale strictly less than the scale of the parent of $m$. The change in branch length $\partial t_m$ is the elongation that occurs with the change in scale $\partial B_m$. The dynamic tree of life is thereby represented as a system of nested random fissions (per Definitions A1, A3, A4 and A5). **(RIGHT)** Illustration showing how the iterated function system represents differential biological form ($\partial B_m$) using scale and differential time elapse ($\partial t_m$) using branch length.

Dropping the requirement of visualization, **Figure 3** gives the scale variables $\beta_i$ for a simulated system of 23 622 paths at $n = 20$ reproductions.

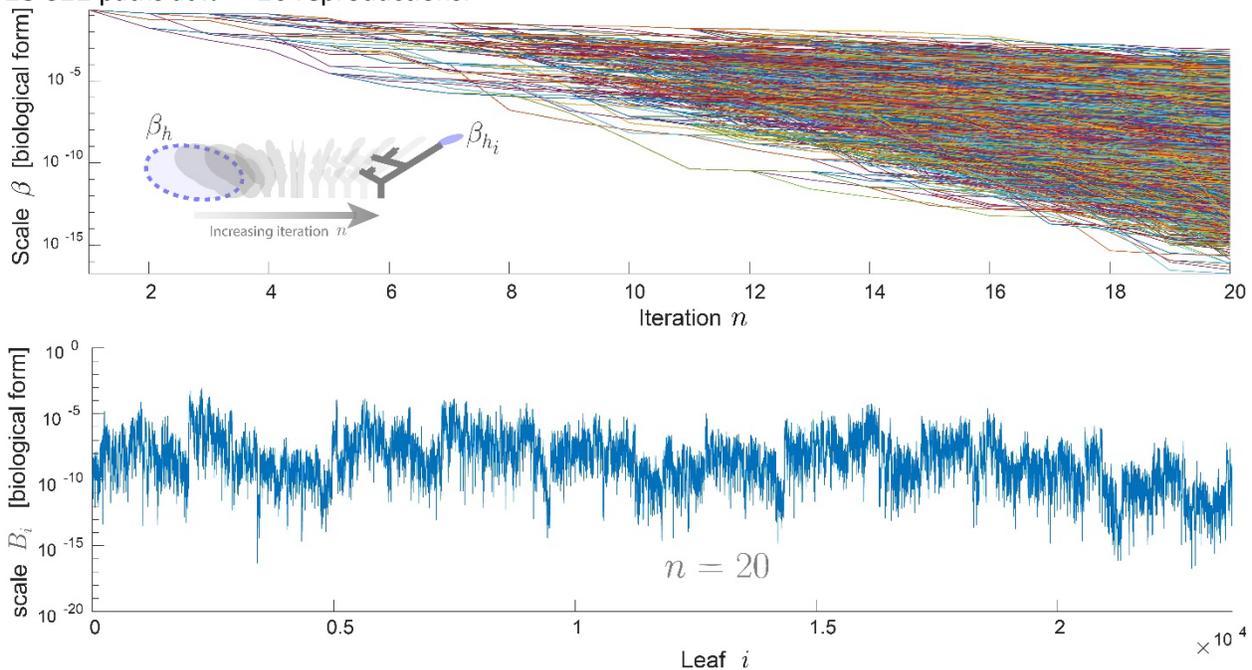

**Figure 3** *Paths in the tree of life.* **(TOP)** The model maps biological form to scale. Here, all 23 622 paths that survive until the 20th iterate are shown for all iterates. **(BOTTOM)** Every path is a monotonically decreasing nested set. As a (Cantor) dust (Theorem 1, Section 3.3), the living are the random variables $B_{i_j}$ taken as the limits over the dead: $B_{i_j} = lim_{n \gg 1} \beta_i$. Here the living are approximated as the leaves at the terminal iterate $n = 20$.

## 4.2 Information theory

We have developed the model of the living tree of life by emphasizing the foundational biological theory it captures (nestedness, duality, randomness). As such, we have left out nearly all analysis of simulation results. Such an analysis is readily achievable using information theory. Ultimately, we will use information measures to link the theory developed in this report to real biological data. However, to keep the present report from becoming excessively lengthy, we leave this development to a future work, and instead offer here a brief preview. The differential **entropy** along any path $\beta_{:l}$ is given by the expression:

Equation 11

$$H\left[B_{:l_m}\right] = -\int_0^{A\left(B_{:l_m}\right)} f_{:l_m}(x) \, log\left(f_{:l_m}(x)\right) dx.$$

Here we have used capital $H$ to represent differential entropy instead of the more common lower case $h$ so as to avoid confusion with nested subscripts.

Since our model implementation assumes uniform probability distributions over monotonically decreasing subintervals of the unit interval (Expression 3), Equation 11 evaluates to the strictly nonpositive expression:

Equation 12

$$H\left[B_{:l_m}\right] = log\left(A\left(B_{:l_m}\right)\right) = \log(\beta_{:l}) \leq 0.$$

Equation 12 says that the entropy of the unborn child is equal to the logarithm of its parent. The entropy along every path for the same system as in Figure 3 is given in **Figure 4**.

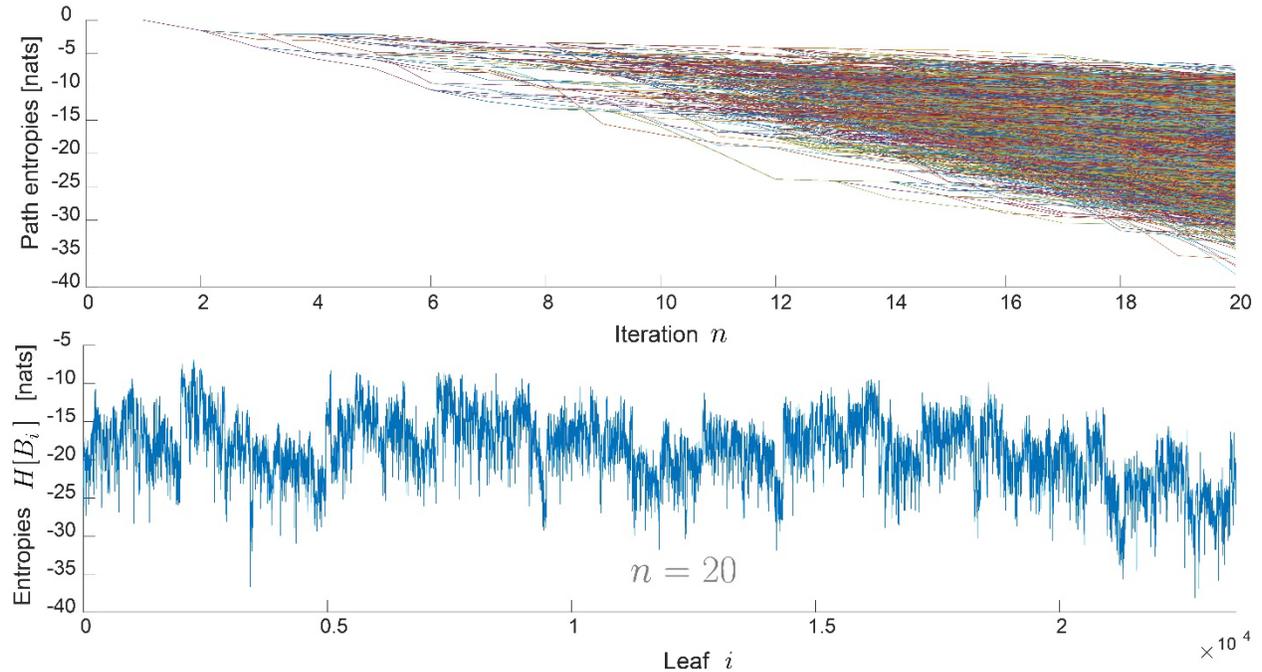

**Figure 4** *Path entropies.* **(TOP)** The entropies along all 23 622 paths that survive until the 20th iterate. Entropy along every path is strictly nonpositive, and strictly increasing in magnitude with increasing iteration. **(BOTTOM)** The entropy of the living is approximated as the large $n$ entropy limit along each unique path. Here that approximation is shown at $n = 20$ reproductions.

## 4.3 The structure is multifractal

It is typically accepted that "fractal" is not a mathematically precise term (Falconer, 2014); and so we might not expect more precision for the term "multifractal". The textbook example of a fractal is a coastline (Mandelbrot, 1982), the length of which depends on the scale of observation. A fractal dimension quantifies how this length scales in relation to the scale of observation. In a monofractal, this relation remains the same across the entire structure so that a single fractal dimension captures the behavior. But in a multifractal, the relation between observation scale and length (or area, or volume, etc.) scales differently in different regions of the structure so that multiple fractal dimensions are required to capture the behavior. Here we show that this is the case for our construction of the tree of life. For a thorough review on multifractals, we refer the reader to (Kantelhardt, 2011).

### 4.3.1 Each path is monofractal

A particular path through the tree of life (denoted here $B$ for simplicity) is a stochastic process that is a function of reproductions $n$:

**Expression 7**

$$\{B(n)\} = \{B_0, \ldots, A(A(\ldots A(B_\alpha) \ldots)), \ldots, A(A(B_\alpha)), A(B_\alpha), B_\alpha\},$$

where $B_0$ is $B(n)$ at $n = 1$, $B_\alpha$ is $B(n)$ at $n = N$, the terminal point, $A(B_\alpha)$ is $B(n)$ at $n = N - 1$, and so on.

Since all $B$ are distributed uniformly for all $n$, with $B(n) < A(B(n)) = B(n-1)$, the following equality of distributions holds:

**Equation 13**

$$B(n) \stackrel{d}{=} n^R B(1), \qquad R < 0,$$

where $R$ is the scale index of the process.

Equation 13 means that the $q$th moment of the process obeys the equality:

**Equation 14**

$$\mathbb{E}[|B(n)|^q] = n^{Rq} \mathbb{E}(B(1)^q), \qquad R < 0,$$

where $\mathbb{E}$ denotes expectation.

The product $Rq$ in Equation 14 is known as the scaling function (Mandelbrot et al., 1997). Since this function is linear, each particular path through the tree of life is uniscaling or monofractal, and therefore a single fractal dimension is sufficient to capture the behavior. However, in general the scale index $R$ will be different for each path. And this is the exact sense in which the tree of life is multifractal: each path scales differently and therefore each path has a different fractal dimension. We next demonstrate this to be the case for simulation results.

### 4.3.2 Each path is a *distinct* monofractal

To show the unique scaling behavior of each path we calculate the generalized Hurst exponents of each path using multifractal detrended fluctuation analysis (Kantelhardt et al., 2002) (**Figure 5**). This method quantifies scaling properties for non-stationary series.

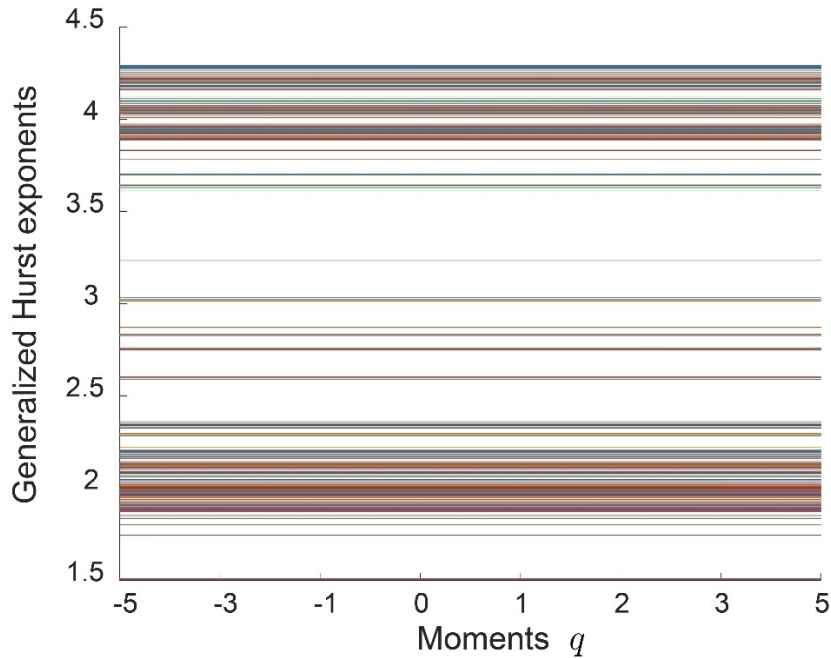

**Figure 5** *The system is multifractal.* Each line in the figure represents a path, with all 23 622 paths shown. The constancy of the generalized Hurst exponent for varying moments $q$ indicates that each path is monofractal as Section 4.3.1 details. That the generalized Hurst exponent is different for each path indicates that each path has a unique scaling behavior, and thus the system as a whole is multifractal.

Figure 5 indicates that each path is indeed a distinct monofractal, and thus the system as a whole is multifractal in the sense that every path is characterized by a different fractal dimension. That these generalized Hurst indices are greater than 1 indicates that each path is dominated by strong trends and long-range correlations. This is expected given each path is a nested unit interval partition (Figure 2).

### 4.3.3 Fractal dimension simulation results

Since we are modeling the tree of life as points and lines, we expect the fractal dimensions to be between zero and one. This is confirmed in **Figure 6**, which gives the fractal dimensions for all paths in the system of 23 622 leaves that survived until iterate 20. For the statistical analysis of determining these fractal dimensions see Appendix C.

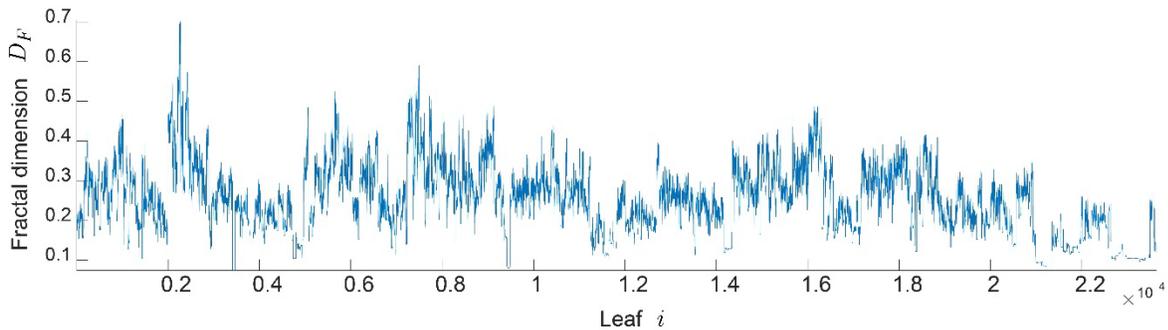

**Figure 6** *Fractal dimensions.* The Fractal dimension is understood to be the correlation dimension. Here the fractal dimensions of all 23 622 leaves at iterate 20 are shown.

## 5 Discussion

We developed a mathematical structure, grounded in fundamental biological principles, that captures the dynamic tree of life by mapping biological form to scale and time to branch length. Using this we showed that a random Cantor-like dust is a mathematical object that captures dynamic phylogenetic space: the living are sequentially realized as biological forms and then made the past. The present is an ever-elusive ephemeral point, continuously "vaporizing" as we "zoom in" but somehow approximated in the existence of the living. Here we discuss some limitations and merits of the model, as well the implications of the multifractal structure, where each branch is a distinct fractal curve.

## 5.1 Modeling finite structures with continuous mathematics

Two limitations to applying Cantor's nested set theorem (Theorem 1) arise due to the finite nature of the tree of life. First, since the number of reproductions is finite, path lengths too are necessarily finite. This means the diameters of the supports of the random variables will never reach zero, as the theorem requires, meaning limits at infinity are never attained. The second limitation lies in the fact that if the theorem did apply to our construction of the tree of life, then our construction would be nowhere dense, and therefore density functions (Equation 4, Section 3.1.2) would not exist.

These two limitations are overcome when we employ the theorem as an approximation of a finite process. We assume that as the pathlength becomes large, there is approximately a single entity at the "end of the path". Under this approximation, density functions do exist, and since the supports of each path contain approximately single entities, they are (approximately) points with topological dimension zero. This "finite but large" approximation is necessary if we are to model the finite structure of the tree of life with continuous mathematics.

### 5.1.2 Convergence in probability

Every path (i.e., every sequence of random variables constituting a lineage as in Figure 3) is monotonically decreasing and bounded below by zero due to the nested nature. In other words, since the support of the child is a subset of the support of the parent on the unit interval (Section 3.1), every path converges in probability to zero as the pathlength goes to infinity (Equation 9). This means that the single thing at the "end" of each path is simply 0. However, since Theorem 1 is employed as an approximation of a finite system, we never arrive at 0. Instead, as $n$ grows large we approximately arrive at 0 so that living entities are approximately points with $D_T = 0$. This approximation allows Definition A1, Appendix A.1 to be met, where a path is characterized as a function from a point with topological dimension zero, to a point with topological dimension zero.

## 5.2 The requirement of sufficient dimensions

One merit of the model of the living tree of life developed herein, is that it is constructed so that biological form $B$ and time $t$ are represented on the same space. Evolution is about a functional dependence of biological form with time: $B = B(t)$. To model this functional dependence, and therefore have a dynamic change in biological form with respect to time $\partial B / \partial t$, form and time must be represented *on the same space*. The model achieves this by mapping biological form to the $n$th-dimensional unit interval as multiplicative, scale-wise nested, sequences of random variables. The existence of scale is a consequence of the nestedness of phylogenetic space, and as such, its importance is widely agreed upon in phylogenetics (Graham et al., 2018; Jablonski, 2000; Whittaker et al., 2001). Using scale to quantify biological form $B$ allows branch length to quantify time $t$ (**Figure 7**).

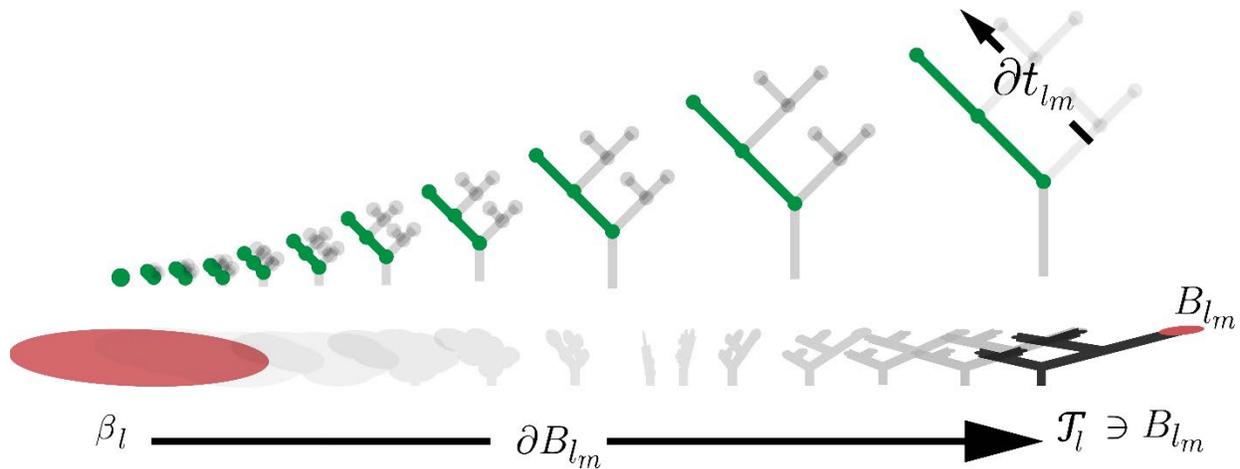

**Figure 7 *Sufficient dimensions.*** The dimension of biological form $B$ is mapped to $[0,1]^n$ as lineage-specific scale paths, where $n$ is equal to the number of reproductions. The figure shows path $B_{l_m}$ from ancestor $l$ to offspring $m$. Time is assigned to branch length. Mapping form to scale and time to branch length allows both to be represented on the same space, thus allowing for a dynamic $\partial B_m / \partial t_m$ for all living $m$ in the system.

### 5.3 Trees within trees

"Trees within trees" is now a phrase in biology at least some thirty years old (Doyle, 1997). The notion of "trees within trees" was implicit in evolutionary models as early as Yule's in 1925 (Kimmel & Axelrod, 2015). But the idea lay mostly dormant until the 1980s when species trees could be constructed from their "nested" gene trees, a practice known as "tree reconciliation" (Avise, 1989; Pamilo & Nei, 1988). In the 1990s it was argued "trees within trees" was a general structure common to biology at all scales (Page & Charleston, 1998). "Trees within trees" is a consequence of the nested principle and the duality principle combined. By utilizing convexity and shifts in topological dimension, our model makes "trees within trees" mathematically explicit. Other important works that make "trees within trees" mathematically explicit include (Blancas et al., 2018; Foutel-Rodier et al., 2021).

### 5.4 Negative entropy

The fact that entropy is a strictly nonpositive quantity in the model deserves a quick word. Negative entropy in the model is due to using continuous random variables with distributions over the unit interval. This is consistent with the notion that biological systems are negative entropy producing processes, which is typically traced back to Schrödinger (1944).

### 5.5 The multifractality of time

The model of the tree of life developed herein generates a Cantor-like dust with a multifractal structure. The notion that phylogeny is fractal has been expounded by many (Chaline et al., 1999; Green, 1991), often with empirical evidence to support the claim (Burlando, 1993; Chaline, 2010; Dubois et al., 1992; Nottale et al., 2002). What we show here is that given phylogeny is nested and dualistic, a fractal structure is expected. But further, when the stochasticity of phylogeny is included, the structure is not just fractal but multifractal; a notion that has also gained attention (Plotnick & Sepkoski, 2001). Multifractality thus follows from accepted theoretical premises of biology: the nested principle, the duality principle, and the random principle. Moreover, the multifractal nature of the living tree of life suggests the multifractality of time. Since branch lengths represent time intervals, and each branch is a distinct fractal, the length of any time interval will depend upon which branch the observation is being made from – in other words, the location of the observer in the system (i.e., its biological form).

### 5.5.1 Time is not an ultrametric on the tree of life

It is typically held that two kinds of metrics can be placed on evolutionary trees: one of biological form and one of time (Bleidorn, 2017; Jantzen et al., 2019; Magallón, 2020). It is also typically claimed either implicitly (Bromham & Penny, 2003; Donoghue & Benton, 2007; A. O. Mooers & Heard, 1997; Raup et al., 1973) or explicitly (Eastman et al., 2013; Marshall, 2008) that time is an ultrametric. The multifractality of the tree of life means the situation is more complicated. Time is only an ultrametric once an observing entity is specified. Independent of specifying an observer, time is multifractal and therefore not a metric. **Figure 8** conceptually illustrates these two different dimensions of the tree of life.

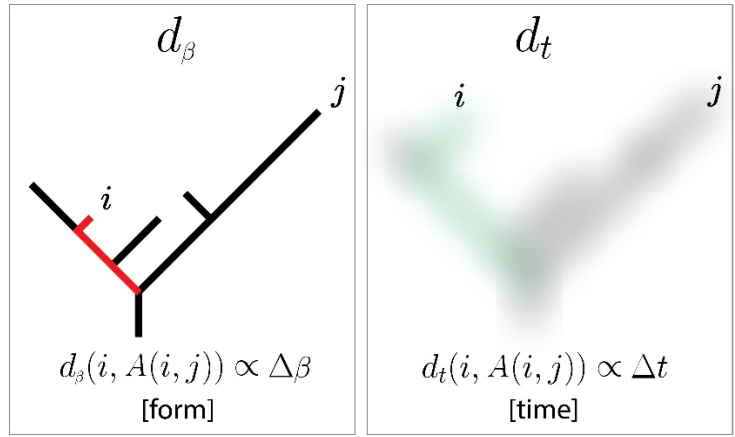

**Figure 8** *Two dimensions of phylogenetic space.* Let all leaves shown on both trees be extant, and let both trees be independent of an observer. An ultrametric is visually identifiable by all extant leaves equidistant from the root. **(LEFT)** $d_\beta$ is not an ultrametric since the amount of change in biological form will vary by lineage. **(MIDDLE)** Independent of specifying an observer, $d_t$ is multifractal and therefore not a metric. The blurred image is meant to represent that there is no fact of the matter as to how long a branch is (i.e., how much time has elapsed) until an observer is specified.

If time is not an ultrametric, but is instead multifractal, then time intervals will depend on the biological form of the observer. **Figure 9** illustrates this functional dependence of time on biological form.

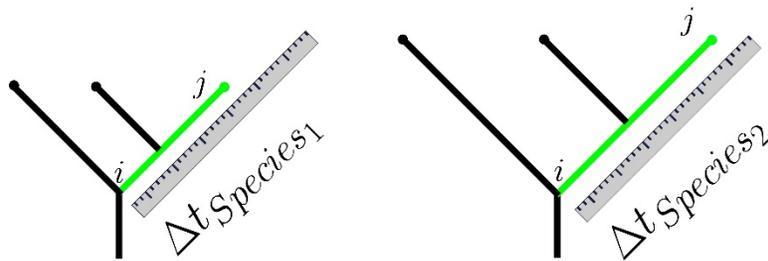

**Figure 9** *Multifractal time.* Two distinct entities from different species each observe the time elapse between events $i$ and $j$ on the tree of life. The multifractality of the tree of life suggests that since the entity from $Species_1$ and the entity from $Species_2$ make the observation from different locations in the tree of life, they will not observe the same time elapse: $\Delta t_{Species_1} \neq \Delta t_{Species_2}$.

### 5.5.2 The biological specificity of defined units

If time intervals (i.e., branch lengths) on the tree of life are distinct fractal curves, then the length of any time interval will vary by the location of the observing entity in the tree of life. Since the location in the tree of life of the observer gives its biological form, this means the length of any time interval will depend on the biological form of the observer. This is the situation depicted in Figure 9. It is a coherent picture of the world, and we can see it as such once we recall that we obtain our SI unit of time, the "second", by *defining* it. This definition is inherently biologically dependent since the definer is a *particular* kind of biological being – in our case, human. Since this definition comes from a particular location in the tree of life, there is no contradiction if two different kinds of entities disagree on durations.

This biological specificity of defined units also means that the multifractal nature of time suggested by this work does not contradict or threaten any established science since all our established science is from the human position only. Once we acknowledge that our units of measurement are definitional, and that the definer is a particular kind of biological entity, the multifractal nature of time is plausible. And more so, it is suggested by established biological theory (the nested dualistic randomness of phylogenetic space).

# 6 Conclusion

Thirty years ago Fontana and Buss found that simple constructive dynamical systems subject to drift alone could exhibit complex biological behavior (1994). They called for a focus on the structure of biological organization as opposed to just the mechanisms. Here we have completely foregone questions of mechanism, and have focused solely on a clearer structural understanding. This clearer structural understanding is obtained from synthesizing three foundational principles of modern biology. The first is the nested principle which follows from the logic of descent-with-modification: a genus leaf *is* a species tree; the species tree *constitutes* the genus leaf. The second is the duality principle, which also follows from the logic of descent-with-modification: biological sets transition between singularities and populations. The third is the random principle, that phylogeny is a stochastic process. These three principles result in a living tree of life that is multifractal. The multifractality arises because each path through the tree of life is a monofractal characterized by a distinct fractal dimension, and hence the tree of life as a whole is multifractal in that it consists of many distinct monofractals.

The model possesses several strengths, which we summarize here. Each of the three fundamental principles of biology are formulated in precise mathematical terms, including relevant terminology (Sections 3.1, 3.2, and Appendix A). The multifractality of the structure generated by the model has been demonstrated mathematically (Section 4.3.1), and validated in simulation results (Section 4.3.2), with fractal dimensions determined for all lineages (Section 4.3.3). Our treatment of convergence is both mathematically responsible (Section 3.3) and biologically realistic (Section 5.1), and the structure generated by the model is well-suited to analysis using information theory (Section 4.2). Additionally, the model successfully captures the dimension of biological form and the dimension of time on the same space (Section 5.2), which stands in distinction to most phylogenetic conceptualizations which represent each on two separate spaces. The model also mathematically captures the widely accepted biological notion of trees-within-trees (Section 5.3). Lastly, the model represents biological systems as negative entropy producing systems, which has been expounded by influential thinkers (Section 5.4).

Establishing the multifractality of the living tree of life is the main result of this work. The length of a fractal curve depends on the location of the observer. In the case of the tree of life, the curves are branch lengths, which give time intervals; and the location of the observer is the location of the observing entity in the tree, which is its biological form. Since branch lengths are distinct fractal curves, it naturally follows that the length of any observed time interval depends on the biological form of the observer.

# Declaration of Competing Interest Statement

The authors declare no known competing personal or financial interests that could have appeared to influence the work reported in this paper.

# CRediT authorship contribution statement

Kevin Hudnall: Conceptualization, Methodology, Software, Writing – Original draft preparation, Visualization, Investigation, Writing – Reviewing and Editing.

Raissa D'Souza: Writing – Reviewing and Editing.

# Acknowledgments

KH would like to thank Tina Jeoh, Shrinivasa Upadhyaya, Nitin Nitin, and Bryan Jenkins, for their early support of this work. KH would also like to thank James Crutchfield for his constructive feedback, and Guram Mikaberidze for many clarifying conversations about mathematical concepts. Finally, we thank three anonymous reviewers whose feedback substantially improved the final version of this paper. This research did not receive any specific grant from funding agencies in the public, commercial, or not-for-profit sectors.

# Appendix A: fundamental definitions

This appendix gives the mathematical definitions for the three foundational principles of biology implemented by the model. $\mathbb{L}^{D_T=0}$ is the set of all possible leaves, and $\mathbb{T}^{D_T=1}$ is the set of all trees (Section 3.1).

## Appendix A.1: Nestedness

The definition of "nested" employed characterizes the "nestedness" of phylogenetic space as a system of convex sequences, which we call "paths".

**Definition A1**

Let $\mathcal{T} \in \mathbb{T}^{D_T=1}$ be a subtree of the tree of life, and let $B_{:i} = A(\mathcal{T}) \in \mathbb{L}^{D_T=0}$ be the ancestor of $\mathcal{T}$. Let $B_j, B_k \in \mathbb{L}^{D_T=0}$ be any two distinct leaves in $\mathcal{T}$, so that $B_j \subset A(\mathcal{T})$ and $B_k \subset A(\mathcal{T})$. Let $f$ be the following map:

$$f: \mathbb{L}^{D_T=0} \longrightarrow \mathbb{L}^{D_T=0}$$

$$f(A(\mathcal{T})) \mapsto B_j, \quad \text{denoted } f_j$$

$$f(A(\mathcal{T})) \mapsto B_k, \quad \text{denoted } f_k$$

with support given as:

$$supp\left(f(A(\mathcal{T}))\right) = \{x \in \mathbb{R} \mid f(A(\mathcal{T}))(x) > 0\}.$$

Then $f_j$ and $f_k$ are both **paths** if the following two conditions hold:

1) $conv(B_j) \subset conv(A(\mathcal{T}))$, and $conv(B_k) \subset conv(A(\mathcal{T}))$,

   but

2) $conv(B_j) \not\subset conv(B_k)$, and $conv(B_k) \not\subset conv(B_j)$,

$$\text{where } conv\left(B_{:i_j}\right) = \left\{ \bigcap_{\alpha=0}^{i} supp\left(f\left(B_{:\alpha:}\right)\right) \right\}, \quad B_{:i} = A(\mathcal{T}),$$

$$\text{and } conv\left(B_{:i_k}\right) = \left\{\bigcap_{\alpha=0}^{i} supp\left(f\left(B_{:\alpha:}\right)\right)\right\}, \quad B_{:i} = A(\mathcal{T}).$$

The subscript notation $(\cdot)_{:\alpha:}$ here indicates that the index $\alpha$ ranges over a nested sequence; and $conv(\cdot)$ is the convex hull.

Paths constitute the tree of life as unique convex hulls. We can now define what it means to be "nested". A structure is **nested** if it contains paths. When $f$ in Definition A1 is taken to be the probability density function (Equation 4, Section 3.1.2), then the definition captures the inherent directionality of nested branching: we inherit from our ancestors not from our siblings, cousins, etc.

The use of convexity has precedent in evolutionary theory: it was an important element of Levin's theory of fitness (Levins, 1962), Williams' approach to determining evolutionary stable strategies in game theory (Williams, 1987), and more recently convex analysis has been used to develop algorithms to determine consensus trees (Chen et al., 2018).

## Appendix A.2: Duality

The set of all possible trees $\mathbb{T}^{D_T=1}$ is concerned with the set of all leaves $\left\{B_{:i_j}, B_{:i_k}, \ldots, B_{:i_\alpha}\right\}$ such that $A\left(B_{:i_j}\right) = A\left(B_{:i_k}\right) = \cdots = A\left(B_{:i_\alpha}\right) = B_{:i}$. For paths, the topological dimension stays constant at zero (Definition A1). But in the transition between leaves and trees, the topological dimension jumps from zero (that of a point) to one (that of a line), so that $\mathcal{T}_{:i} \in \mathbb{T}^{D_T=1}$ is equal to the set of respective distances (generically denoted $d$) between the leaves and the ancestor:

**Definition A2**

$$\mathcal{T}_{:i} \equiv \left\{d\left(B_{:i_j}, A\left(B_{:i_j}\right)\right), d\left(B_{:i_k}, A\left(B_{:i_k}\right)\right), \ldots, d\left(B_{:i_\alpha}, A\left(B_{:i_\alpha}\right)\right)\right\},$$

$$\text{with } A\left(B_{:i_j}\right) = A\left(B_{:i_k}\right) = \cdots = A\left(B_{:i_\alpha}\right) = B_{:i}.$$

Though the topological dimension changes, the basic convexity still holds: a species tree is "contained within" a genus leaf in the sense that every member of the tree is "contained within" the ancestor of the tree. Using this convexity, we can define these "nested fissions", and thereby arrive at a definition of "duality".

**Definition A3**

Let $B_{:i} \in \mathbb{L}^{D_T=0}$ be a leaf in the tree of life. Let each member of the set $\left\{B_{:i_j}, B_{:i_k}, \ldots, B_{:i_\alpha}\right\}$ also be a leaf in the tree of life with $A\left(B_{:i_j}\right) = A\left(B_{:i_k}\right) = \cdots = A\left(B_{:i_\alpha}\right) = B_{:i}$, so that $\mathcal{T}_{:i} \in \mathbb{T}^{D_T=1}$ is the tree generated by $B_{:i}$:

$$\mathcal{T}_{:i} = \left\{d\left(B_{:i_j}, A\left(B_{:i_j}\right)\right), d\left(B_{:i_k}, A\left(B_{:i_k}\right)\right), \ldots, d\left(B_{:i_\alpha}, A\left(B_{:i_\alpha}\right)\right)\right\},$$

$$\text{and } A(\mathcal{T}_{:i}) = B_{:i}.$$

Let $g$ be the following map:

$$g: \mathbb{L}^{D_T=0} \to \mathbb{T}^{D_T=1}$$

$$g\left(A(\mathcal{T}_{:i})\right) \mapsto \mathcal{T}_{:i},$$

with support given as:

$$supp\left(g\left(A(\mathcal{T}_{:i})\right)\right) = \left\{x \in \mathbb{R} \,\middle|\, g\left(A(\mathcal{T}_{:i})(x)\right) > 0\right\}.$$

Then $g$ is a **convex fission** (or just **fission** for simplicity) if for each member $d\left(B_{:i_\alpha}, A\left(B_{:i_\alpha}\right)\right)$ of the set $\mathcal{T}_{:i}$, $f_{:i_\alpha}$ is a path.

We can now define what it means for a system to be "dualistic". A system is **dualistic** if it contains fissions.

In the iterated function system, $g$ is the branching process (i.e., the function that is iterated) (Equation 6, Section 3.2). The fissions are biological transitions, where the topological dimension shifts from 0 to 1 as a random variable is realized (i.e., as "the coin lands" and the future living become the living while the living become the dead).

### Appendix A.3: Life and death

The definitions of life and death given here capture the notion expressed by Von Bertalanffy in 1942 that "living forms are not *in being*, they are *happening*" (Ganti, 1979). A biological entity is **living/extant** ("happening") if there exists at least one kind of biological form it has a nonzero probability of becoming. Such an entity is in the **living tree of life**:

**Definition A4**

$$\mathbb{T}^{\text{alive}} = \left\{\text{all } m \text{ for which there exists an } x \text{ such that } Pr(B_{:m} \leq x) \neq \{0,1\}\right\}.$$

And likewise we can define a dead tree of life. A biological entity is **dead/extinct** ("not happening") if the probability of it being any other form is 0 or 1. Such an entity is in the **dead tree of life**:

**Definition A5**

$$\mathbb{T}^{\text{dead}} = \left\{\text{all } m \text{ for which } Pr(B_{:m} \leq x) = 0 \text{ or } 1 \text{ for all } x\right\}.$$

The condition in Definition A.4 that $Pr(B_{:m} \leq x)$ cannot equal one is due to the following. Suppose $\beta_{:m}$ is dead. Then $Pr(B_{:m} = \beta_{:m}) = 1$. The "coin has already been flipped"; $B_{:m}$ has realized its form as $\beta_{:m}$; so with probability one, $B_{:m}$ is equal to $\beta_{:m}$, and therefore has zero probability of being any other form.

## Appendix B: Illustrative example of the iterated function system

Here, given in **Figure B1**, is a simple example of the iterated function system starting at iterate zero and going through three iterations. Iterate zero represents the first ancestor of all life. There are no ratio lists or function lists. Instead, iterate zero corresponds to the initial condition: $Pr(B_0 = \beta_0 = 1) = 1$.

| Iterate | Illustration | Ratio lists | Function lists |
|---|---|---|---|
| 0 | $\bullet\ B_0$ | N/A | N/A |
| 1 | $scale(\mathcal{T}_0) \sim U(0, \beta_0)$ <br> $B_{0_1}\ B_{0_2}\ B_{0_3}$ | $\{\beta_0\}$ | $\{T_0^0\}$ |
| 2 | $scale(\mathcal{T}_{0_1}) \sim U(0,\beta_{0_1})$, $scale(\mathcal{T}_{0_2}) \sim U(0,\beta_{0_2})$, $scale(\mathcal{T}_{0_3}) \sim U(0,\beta_{0_3})$ <br> $B_{0_{1_1}}\ B_{0_{1_2}}\ B_{0_{2_1}}\ B_{0_{2_2}}\ B_{0_{2_3}}\ B_{0_{3_1}}\ B_{0_{3_2}}$ | $\{\beta_{0_1}, \beta_{0_2}, \beta_{0_3}\}$ | $\{T_{0_1}^1, T_{0_2}^1, T_{0_3}^1\}$ |
| 3 | $scale(\mathcal{T}_{0_{1_1}}) \sim U(0,\beta_{0_{1_1}})$, $scale(\mathcal{T}_{0_{1_2}}) \sim U(0,\beta_{0_{1_2}})$, $scale(\mathcal{T}_{0_{2_1}}) \sim U(0,\beta_{0_{2_1}})$, $scale(\mathcal{T}_{0_{2_2}}) \sim U(0,\beta_{0_{2_2}})$, $scale(\mathcal{T}_{0_{2_3}}) \sim U(0,\beta_{0_{2_3}})$, $scale(\mathcal{T}_{0_{3_1}}) \sim U(0,\beta_{0_{3_1}})$, $scale(\mathcal{T}_{0_{3_2}}) \sim U(0,\beta_{0_{3_2}})$ | $\{\beta_{0_{1_1}}, \beta_{0_{1_2}}\}$ <br> $\{\beta_{0_{2_1}}, \beta_{0_{2_2}}, \beta_{0_{2_3}}\}$ <br> $\{\beta_{0_{3_1}}, \beta_{0_{3_2}}\}$ | $\{T_{0_{1_1}}^2, T_{0_{1_2}}^2\}$ <br> $\{T_{0_{2_1}}^2, T_{0_{2_2}}^2, T_{0_{2_3}}^2\}$ <br> $\{T_{0_{3_1}}^2, T_{0_{3_2}}^2\}$ |

**Figure B1** *Example of the iterated function system.* Three hypothetical iterations of the iterated function system are shown, resulting in a system with a single subtree at iterate 1, three subtrees at iterate 2, and seven subtrees at iterate 3.

Moving to iterate 1, there is a single ratio list and it contains a single member $\beta_0$. Likewise, there is a single function list with a single member. The function generates the first tree $\mathcal{T}_0$, where the scale of $\mathcal{T}_0$ is uniformly distributed from 0 to $\beta_0$. The members of $\mathcal{T}_0$ are $\{B_{0_1}, B_{0_2}, B_{0_3}\}$. At this point, they all have a nonzero probability of fission, and so they are represented as capital betas (i.e., living) instead of lower case betas (i.e., dead).

In iterate 2, each member of $\mathcal{T}_0$ has now gone through fission, so the ratio list (of which there is still just one) consists of the members of $\mathcal{T}_0$ now represented as lower case betas: $\{\beta_{0_1}, \beta_{0_2}, \beta_{0_3}\}$. To this ratio list there corresponds the function list $\{T_{0_1}^1, T_{0_2}^1, T_{0_3}^1\}$ which generates the subtrees $\{\mathcal{T}_{0_1}, \mathcal{T}_{0_2}, \mathcal{T}_{0_3}\}$, where the scale of each random tree is uniformly distributed between 0 and its corresponding member of the ratio list.

Moving to iterate 3, there are now three separate ratio lists, one for each member of the ratio list in iterate 2. And likewise there are three corresponding function lists. Again, each member of each function list generates a random tree where the scale of the random tree is distributed uniformly between 0 and the member of its corresponding ratio list. Note that since throughout the iterative process, each leaf of a subtree is at the same scale as every other leaf in the subtree, all the members of a particular ratio list are equal, but the values between ratio lists are not. So in iterate 3 for example, $\beta_{0_{2_1}} = \beta_{0_{2_2}} = \beta_{0_{2_3}}$; but $\beta_{0_{2_1}} \neq \beta_{0_{1_1}}$, $\beta_{0_{3_1}} \neq \beta_{0_{2_1}}$, and so on.

# Appendix C: Statistical analysis for correlation dimensions

To obtain $D_F$ we generate an array $\boldsymbol{\varepsilon}$ of 12 000 logarithmically descending values between 1 and 0, but not including 1 and 0. For the particular path under consideration, we step through each value of $\boldsymbol{\varepsilon}$ and determine the number of pairs of points whose separation distance is less than that value of $\boldsymbol{\varepsilon}$. This value is then divided by the square of the total number of points in the path, so that the process generates the array $\boldsymbol{C(\varepsilon)}$. Once we arrive at a value of $\boldsymbol{\varepsilon}$ for which the corresponding value of $\boldsymbol{C(\varepsilon)}$ is zero, we stop and exclude that value of $\boldsymbol{\varepsilon}$ and $\boldsymbol{C(\varepsilon)}$. A linear regression is performed on $log(\boldsymbol{\varepsilon})$ versus $log(\boldsymbol{C(\varepsilon)})$, and $D_F$ is taken to be the slope of this regression line.

This method of determining the correlation dimension yields reasonably good statistics. **Figure C1 A** presents such a graph for a randomly chosen path from the system of 23 622 paths, along with the coefficient of determination $R^2$. **Figure C1 B** gives a histogram of $R^2$ values for all the paths in the system of 23 622 paths that survived until iterate 20; and **Figure C1 C** gives the average $R^2$ value at each iteration. We see that the method becomes reliable at about iteration 7, after which neither the average $R^2$ value nor the standard deviation changes significantly.

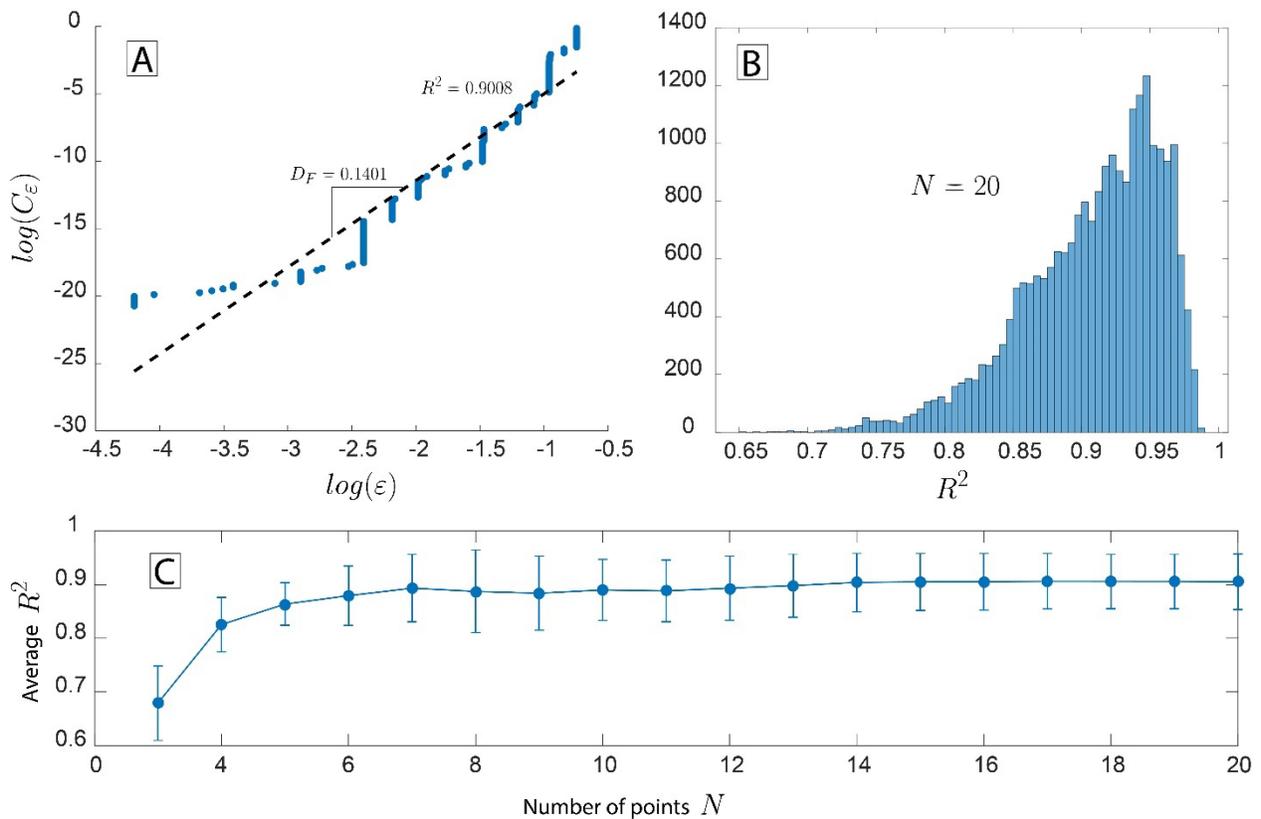

**Figure C1** *Statistics for determining the correlation dimension.* **(A)** For a particular path through the tree of life, the correlation dimension $D_F$ is determined by regressing the log of the correlation integral $C(\varepsilon)$ against the log of the separation distances $\varepsilon$. $D_F$ is then the slope of the regression line. Here a particular path was randomly chosen and taken up to the 20th iterate so that the total number of points is 20. There are 56 values for the separation distance descending logarithmically from 1 to $7.9^{-12}$. **(B)** A histogram for the coefficients of determination $R^2$ in the population at the 20th iterate. The average $R^2$ value at iterate 20 is $0.9006 \pm 0.0542$. **(C)** Average $R^2$ values at each iterate with error bars giving standard deviations. Very minimal change occurs after iterate 7, with essentially no change after iterate 14.

## Data availability

The data used in this paper as well as fully executable files can be found in the git repository https://github.com/KevinAndrewHudnall/the-living-tree-of-life.git.

## References

Ane, C., Ho, L. S. T., & Roch, S. (2017). Phase transition on the convergence rate of parameter estimation under an Ornstein-Uhlenbeck diffusion on a tree. *Journal of Mathematical Biology*, *74*(1–2), 355–385. https://doi.org/10.1007/s00285-016-1029-x

Archibald, J. D. (2009). Edward Hitchcock's pre-Darwinian (1840) "tree of life." *Journal of the History of Biology*, *42*(3), 561–592. https://doi.org/10.1007/s10739-008-9163-y

Avise, J. C. (1989). Gene Trees and Organismal Histories: A Phylogenetic Approach to Population Biology. *Evolution*, *43*(6), 1192–1208. https://doi.org/10.1111/j.1558-5646.1989.tb02568.x

Bellman, R., & Harris, T. (1949). On Age-Dependent Stochastic Branching Processes. *Annals of Mathematical Statistics*, *20*(1), 139–139.

Blancas, A., Duchamps, J.-J., Lambert, A., & Siri-Jégousse, A. (2018). Trees within trees: Simple nested coalescents. *Electronic Journal of Probability*, *23*(none), 1–27. https://doi.org/10.1214/18-EJP219

Bleidorn, C. (2017). *Phylogenomics: An Introduction* (1st ed. 2017 edition). Springer.

Bromham, L., & Penny, D. (2003). The modern molecular clock. *Nature Reviews Genetics*, *4*(3), 216–224. https://doi.org/10.1038/nrg1020

Burlando, B. (1993). The Fractal Geometry of Evolution. *Journal of Theoretical Biology*, *163*(2), 161–172. https://doi.org/10.1006/jtbi.1993.1114

Chaline, J. (2010). Does Species Evolution Follow Scale Laws? First Applications of the Scale Relativity Theory to Fossil and Living-beings. *Foundations of Science*, *15*(3), 279–302. https://doi.org/10.1007/s10699-010-9180-0

Chaline, J., Nottale, L., & Grou, P. (1999). Is the evolutionary tree a fractal structure? *Comptes Rendus De L Academie Des Sciences Serie Ii Fascicule a-Sciences*, *328*(11), 717–726. https://doi.org/10.1016/S1251-8050(99)80162-1

Chen, S., Shi, P., Lim, C.-C., & Lu, Z. (2018). A consensus algorithm in CAT(0) space and its application to distributed fusion of phylogenetic trees. *Journal of Mathematical Analysis and Applications*, *459*(2), 1149–1159. https://doi.org/10.1016/j.jmaa.2017.11.028

Chou, T., & Greenman, C. (2016). A Hierarchical Kinetic Theory of Birth, Death and Fission in Age-Structured Interacting Populations. *Journal of Statistical Physics*. https://doi.org/10.1007/s10955-016-1524-x

Cohn, D. L. (2013). *Measure Theory: Second Edition* (2nd ed. 2013 edition). Birkhäuser.

Czuppon, P., & Gokhale, C. S. (2018). Disentangling eco-evolutionary effects on trait fixation. *Theoretical Population Biology*, *124*, 93–107. https://doi.org/10.1016/j.tpb.2018.10.002

Darwin, C. (1859). *On the Origin of Species by Means of Natural Selection, or the Preservation of Favoured Races in the Struggle for Life*. John Murray.

Dobzhansky, T. (1973). Nothing in Biology Makes Sense except in the Light of Evolution. *The American Biology Teacher*, *35*(3), 125–129. https://doi.org/10.2307/4444260

Donoghue, P. C. J., & Benton, M. J. (2007). Rocks and clocks: Calibrating the Tree of Life using fossils and molecules. *Trends in Ecology & Evolution*, *22*(8), 424–431. https://doi.org/10.1016/j.tree.2007.05.005

Doulcier, G., Lambert, A., De Monte, S., & Rainey, P. B. (2020). Eco-evolutionary dynamics of nested Darwinian populations and the emergence of community-level heredity. *eLife*, *9*, e53433. https://doi.org/10.7554/eLife.53433

Doyle, J. J. (1997). Trees within trees: Genes and species, molecules and morphology. *Systematic Biology*, *46*(3), 537–553. https://doi.org/10.2307/2413695

Dubois, J., Chaline, J., & Brunetlecomte, P. (1992). Speciation, Extinction and Strange Attractors. *Comptes Rendus De L Academie Des Sciences Serie Ii*, *315*(13), 1827–1833.

Eastman, J. M., Harmon, L. J., & Tank, D. C. (2013). Congruification: Support for time scaling large phylogenetic trees. *Methods in Ecology and Evolution*, *4*(7), 688–691. https://doi.org/10.1111/2041-210X.12051

Eigen, M. (2000). Natural selection: A phase transition? *Biophysical Chemistry*, *85*(2), 101–123. https://doi.org/10.1016/S0301-4622(00)00122-8

Falconer, K. (2014). *Fractal Geometry: Mathematical Foundations and Applications* (3 edition). Wiley.

Fontana, W., & Buss, L. W. (1994). "The arrival of the fittest": Toward a theory of biological organization. *Bulletin of Mathematical Biology*, *56*(1), 1–64. https://doi.org/10.1007/BF02458289

Fontana, W., & Schuster, P. (1998). Continuity in evolution: On the nature of transitions. *Science*, *280*(5368), 1451–1455. https://doi.org/10.1126/science.280.5368.1451

Foutel-Rodier, F., Lambert, A., & Schertzer, E. (2021). Exchangeable coalescents, ultrametric spaces, nested interval-partitions: A unifying approach. *The Annals of Applied Probability*, *31*(5), 2046–2090. https://doi.org/10.1214/20-AAP1641

Ganti, T. (1979). *A theory of biochemical supersystems and its application to problems of natural and artificial biogenesis*. University Park Press.

Gernhard, T., Hartmann, K., & Steel, M. (2008). Stochastic properties of generalised Yule models, with biodiversity applications. *Journal of Mathematical Biology*, *57*(5), 713–735. https://doi.org/10.1007/s00285-008-0186-y

Graham, C. H., Storch, D., & Machac, A. (2018). Phylogenetic scale in ecology and evolution. *Global Ecology and Biogeography*, *27*(2), 175–187. https://doi.org/10.1111/geb.12686

Grassberger, P., & Procaccia, I. (1983). Measuring the strangeness of strange attractors. *Physica D: Nonlinear Phenomena*, *9*(1), 189–208. https://doi.org/10.1016/0167-2789(83)90298-1

Green, D. M. (1991). Chaos, fractals and nonlinear dynamics in evolution and phylogeny. *Trends in Ecology & Evolution*, *6*(10), 333–337. https://doi.org/10.1016/0169-5347(91)90042-V

Grey, D., Hutson, V., & Szathmáry, Eö. (1995). A Re-Examination of the Stochastic Corrector Model. *Proceedings: Biological Sciences*, *262*(1363), 29–35.

Hallinan, N. (2012). The generalized time variable reconstructed birth-death process. *Journal of Theoretical Biology*, *300*, 265–276. https://doi.org/10.1016/j.jtbi.2012.01.041

Jablonka, E. (1994). Inheritance Systems and the Evolution of New Levels of Individuality. *Journal of Theoretical Biology*, *170*(3), 301–309. https://doi.org/10.1006/jtbi.1994.1191

Jablonski, D. (2000). Micro- and macroevolution: Scale and hierarchy in evolutionary biology and paleobiology. *Paleobiology*, *26*(4), 15–52. https://doi.org/10.1666/0094-8373(2000)26[15:MAMSAH]2.0.CO;2

Jantzen, J. R., Whitten, W. M., Neubig, K. M., Majure, L. C., Soltis, D. E., & Soltis, P. S. (2019). Effects of taxon sampling and tree reconstruction methods on phylodiversity metrics. *Ecology and Evolution*, *9*(17), 9479–9499. https://doi.org/10.1002/ece3.5425

Kantelhardt, J. W. (2011). Fractal and Multifractal Time Series. In R. A. Meyers (Ed.), *Mathematics of Complexity and Dynamical Systems* (pp. 463–487). Springer. https://doi.org/10.1007/978-1-4614-1806-1_30

Kantelhardt, J. W., Zschiegner, S. A., Koscielny-Bunde, E., Havlin, S., Bunde, A., & Stanley, H. E. (2002). Multifractal detrended fluctuation analysis of nonstationary time series. *Physica A: Statistical Mechanics and Its Applications*, *316*(1), 87–114. https://doi.org/10.1016/S0378-4371(02)01383-3

Kimmel, M., & Axelrod, D. E. (2015). *Branching Processes in Biology* (2nd ed. 2015 edition). Springer.

Levins, R. (1962). Theory of Fitness in a Heterogeneous Environment .1. Fitness Set and Adaptive Function. *American Naturalist*, *96*(891), 361-. https://doi.org/10.1086/282245

Loeffler, M., & Grossmann, B. (1991). A stochastic branching model with formation of subunits applied to the growth of intestinal crypts. *Journal of Theoretical Biology*, *150*(2), 175–191. https://doi.org/10.1016/s0022-5193(05)80330-3

Magallón, S. (2020). Principles of Molecular Dating. In S. Y. W. Ho (Ed.), *The Molecular Evolutionary Clock: Theory and Practice*. Springer International Publishing. https://doi.org/10.1007/978-3-030-60181-2

Mandelbrot, B. B. (1982). *The Fractal Geometry of Nature* (2nd prt. edition). Times Books.

Mandelbrot, B. B., Fisher, A. J., & Calvet, L. E. (1997). *A Multifractal Model of Asset Returns* (SSRN Scholarly Paper 78588). https://papers.ssrn.com/abstract=78588

Marshall, C. R. (2008). A Simple Method for Bracketing Absolute Divergence Times on Molecular Phylogenies Using Multiple Fossil Calibration Points. *The American Naturalist*, *171*(6), 726–742. https://doi.org/10.1086/587523

McShea, D. W. (2001). The minor transitions in hierarchical evolution and the question of a directional bias. *Journal of Evolutionary Biology*, *14*(3), 502–518. https://doi.org/10.1046/j.1420-9101.2001.00283.x

Mooers, A., Gascuel, O., Stadler, T., Li, H., & Steel, M. (2012). Branch Lengths on Birth-Death Trees and the Expected Loss of Phylogenetic Diversity. *Systematic Biology*, *61*(2), 195–203. https://doi.org/10.1093/sysbio/syr090

Mooers, A. O., & Heard, S. B. (1997). Inferring Evolutionary Process from Phylogenetic Tree Shape. *The Quarterly Review of Biology*, *72*(1), 31–54. https://doi.org/10.1086/419657

Nottale, L., Chaline, J., & Grou, P. (2002). On the Fractal Structure of Evolutionary Trees. In G. A. Losa, D. Merlini, T. F. Nonnenmacher, & E. R. Weibel (Eds.), *Fractals in Biology and Medicine* (pp. 247–258). Birkhäuser Basel.

Page, R. D. M., & Charleston, M. A. (1998). Trees within trees: Phylogeny and historical associations. *Trends in Ecology & Evolution*, *13*(9), 356–359. https://doi.org/10.1016/S0169-5347(98)01438-4

Pamilo, P., & Nei, M. (1988). Relationships between gene trees and species trees. *Molecular Biology and Evolution*, *5*(5), 568–583. https://doi.org/10.1093/oxfordjournals.molbev.a040517

Plotnick, R. E., & Sepkoski, J. J. (2001). A Multiplicative Multifractal Model for Originations and Extinctions. *Paleobiology*, *27*(1), 126–139.

Ponisio, L. C., Valdovinos, F. S., Allhoff, K. T., Gaiarsa, M. P., Barner, A., Guimaraes Jr, P. R., Hembry, D. H., Morrisong, B., & Gillespie, R. (2019). A Network Perspective for Community Assembly. *Frontiers in Ecology and Evolution*, *7*, 103. https://doi.org/10.3389/fevo.2019.00103

Raup, D. M., Gould, S. J., Schopf, T. J. M., & Simberloff, D. S. (1973). Stochastic Models of Phylogeny and the Evolution of Diversity. *The Journal of Geology*, *81*(5), 525–542. https://www.jstor.org/stable/30060095

Schrodinger, E. (1944). *What is Life?: With Mind and Matter and Autobiographical Sketches*. Cambridge University Press.

Semple, C. (2016). Phylogenetic Networks with Every Embedded Phylogenetic Tree a Base Tree. *Bulletin of Mathematical Biology*, *78*(1), 132–137. https://doi.org/10.1007/s11538-015-0132-2

Shelton, D. E., & Michod, R. E. (2014). Group Selection and Group Adaptation During a Major Evolutionary Transition: Insights from the Evolution of Multicellularity in the Volvocine Algae. *Biological Theory*, *4*(9), 452–469. https://doi.org/10.1007/s13752-014-0159-x

Szathmáry, E., & Smith, J. M. (1995). The major evolutionary transitions. *Nature*, *374*(6519), 227–232. https://doi.org/10.1038/374227a0

Whittaker, R. J., Willis, K. J., & Field, R. (2001). Scale and species richness: Towards a general, hierarchical theory of species diversity. *Journal of Biogeography*, *28*(4), 453–470. https://doi.org/10.1046/j.1365-2699.2001.00563.x

Williams, H. P. (1987). Evolution, games theory and polyhedra. *Journal of Mathematical Biology*, *25*(4), 393–409. https://doi.org/10.1007/BF00277164

Yule, G. U. (1925). A Mathematical Theory of Evolution, Based on the Conclusions of Dr. J. C. Willis, F.R.S. *Philosophical Transactions of the Royal Society of London. Series B, Containing Papers of a Biological Character*, *21*, 21–87.

## Silhouette credits

All silhouettes were accessed at phylopic.org.

*Rocio octofasciata*. By Camilo Julián-Caballero. License found at https://creativecommons.org/licenses/by/3.0/.

*Cheilinus chlorourus*. By Kent Sorgon. License found at https://creativecommons.org/licenses/by-sa/3.0/.

*Perca flavescens*. By NOAA Great Lakes Environmental Research Laboratory (illustration) and Timothy J. Bartley. License found at https://creativecommons.org/licenses/by-sa/3.0/.

*Sander lucioperca*. By Carlos Cano-Barbacil.

*Australoheros facetus.* By Carlos Cano-Barbacil.

*Lepomis macrochirus*. Uncredited

*Umbrina roncador*. By Nick Schooler.